\documentclass[fleqn,floatfix,twocolumn,showpacs,preprintnumbers,amsmath,amssymb,prl]{revtex4}
\usepackage{graphicx}
\usepackage{dcolumn}
\usepackage{bm}

\begin{document}

\preprint{Submitted to Condmat}

\title{The Kondo Screening Cloud around a Quantum Dot: Large-scale numerical Results}

\author{Erik~S.~S\o rensen}%
\affiliation{Department of Physics and Astronomy, McMaster University, Hamilton,
ON, L8S 4M1 Canada }
\author{Ian Affleck}
\affiliation{Department of Physics and Astronomy, University of British Columbia, Vancouver, British Columbia, Canada V6T 1Z1}

\date{\today}

\begin{abstract}
Measurements of the persistent current 
in a ring containing a quantum dot would afford a unique opportunity 
to finally detect the elusive Kondo screening cloud. 
We present the 
first large-scale numerical results on this controversial subject using 
exact diagonalization and density matrix renormalization group (RG). 
These extremely challenging numerical calculations 
confirm  RG arguments for weak to strong coupling crossover 
with varying ring length and give results on 
the universal scaling functions. 
We also study, analytically and numerically, the important and
surprising effects of particle-hole symmetry breaking.
\end{abstract}

\maketitle
The screening of an impurity spin by conduction electrons, the Kondo effect, 
is believed by many to be associated with the formation of a 
``screening cloud'' around the impurity with a size $\xi_K=v_F/T_K$ 
where $v_F$ is the Fermi velocity and $T_K$ is the Kondo temperature, 
the characteristic energy scale associated with this screening~\cite{Nozieres}. 
This fundamental length scale, which has been called the ``holy grail" of Kondo research~\cite{KouwenGlazman},
and which from the above estimate 
can be as large as 1 micron in typical situations, has never 
been observed experimentally and has sometimes been questioned 
theoretically. Although traditionally associated with 
dilute impurity spins in metals, the Kondo effect has been 
observed more recently in nano-structures\cite{Goldhaber,Nygaard,Manoharan,Wiel}.  The electron number  
on semi-conductor quantum dots, weakly coupled to leads, 
can be varied in single steps with a gate voltage. When 
this electron number is odd, the quantum dot generally 
has a spin of 1/2 and  can act as a 
Kondo impurity, screened by electrons in the leads. By 
attaching the dot to quantum wires of length $L$, the 
Kondo screening cloud size could be measured from 
the dependence of conductance properties on $L$\cite{SAprl,SAprb} A particularly
simple case is when the dot is attached to a single quantum 
wire which forms a closed ring. Two simple tight-binding models 
have been considered corresponding to an ``embedded'' or 
``side-coupled'' quantum dot   (EQD or SCQD).
Suppressing electron spin indices, the corresponding Hamiltonians are:
\begin{equation}
H_{\rm EQD}=-t\sum_{j=1}^{L-2}\left(c^{\dagger}_jc_{j+1}+{\rm h.c.}\right)+H_K,
\label{eq:H_1}
\end{equation}
where
$
H_K=J_K{\vec S}\cdot (c_1^{\dagger}+c_{L-1}^{\dagger})
\frac{\vec\sigma}{2}(c_1+c_{L-1}), 
$
and 
\begin{equation}
H_{\rm SCQD}=-t\sum_{j=0}^{L-1}\left(c^{\dagger}_jc_{j+1}+{\rm h.c.}\right)+H_K,
\ \  (c_L\equiv c_0)\label{eq:H_2}
\end{equation}
where $H_K=J_K{\vec S}\cdot c_0^{\dagger}
\frac{\vec\sigma}{2}c_0$, and
$c_{j\sigma}$ annihilates an electron at site $j$ of spin $\sigma$; 
$S^a$ are S=1/2 spin operators. 
We generally set $t=1$ in what follows. 
A magnetic flux is added to the model by adding appropriate 
phases to the hopping terms and the resulting 
persistent current, $j$, is measured. 

The EQD must have $j=0$
when the Kondo coupling, $J_K=0$, since then 
the sites 1 and (-1) are decoupled from each other. On the 
other hand, the SCQD has a large $j$
when $J_K=0$ since the 
Hamiltonian then reduces to that of a free ring with periodic 
boundary conditions. When the Kondo 
coupling becomes large ($J_K\gg t$) essentially the inverse behavior occurs, 
due to the formation of a Kondo screening cloud.  In the strong coupling 
limit of the EQD 
the screening electron goes into the symmetric orbital on sites 
1 and -1. This allows  resonant transmission through the anti-symmetric 
orbital on these sites and the ideal saw-tooth like $j$ of 
a free ring occurs, when the system is at 1/2-filling. 
On the 
other hand, for the SCQD, the screening electron sits at site-0 
and completely blocks all current flow in the strong coupling limit. 
The more interesting question is the behavior of $j$ in these models 
for small $J_K/t$ at large length, $L$. The Kondo length scale 
grows exponentially as $J_K/t\to 0$. The persistent current 
was predicted~\cite{SAprl,SAprb} to be given by universal scaling functions of 
$\xi_K/L$ and the dimensionless magnetic flux through the 
ring, $\alpha \equiv e\Phi /c$:
\begin{equation}
jL/(ev_F)=f(\xi_K/L, \alpha ), \ \ L,\xi_K\gg a,
\end{equation}
with $a$ the lattice constant.
The crossover functions, $f$, are very different for the EQD and 
SCQD and also depend on the parity of the electron number in the ring, $N$. 
However, they are otherwise expected to be universal in 
the small $J_K$ limit, not depending on details of the 
dispersion relation, electron density, 
the range of the Kondo interaction, etc. 
In previous work by one of us and P.~Simon~\cite{SAprl,SAprb} 
it was argued that, in the limit $L\gg\xi_K$, strong coupling 
behavior occurs, since the effective Kondo coupling 
is expected to become large at large length scales. Thus 
it was predicted that $f(\xi_K/L,\alpha )$ approaches 
the value for an ideal ring at $\xi_K/L\to 0$ for 
the EQD but approaches zero for the SCQD, for either parity of $N$. 
The former prediction is in disagreement with other studies~\cite{Kang,Ferrari}
where $jL$ for the EQD was predicted to have very different $L$-dependence, whereas the 
SCQD prediction is in stark contradiction with the conclusions of other
authors~\cite{Eckle,Cho} who predicted that $j$ attains the value for an ideal
ring, at $\xi_K/L\to 0$. 
However, the results of Refs~\onlinecite{SAprl,SAprb}
are in agreement with the theoretical, numerical (up to $L=8$) and
mean field results of Refs.~\onlinecite{Hu,Aligia,Cornaglia,Hallberg}.
Other theoretical studies include 
Refs.~\onlinecite{ZvyaginJLT,Zvyaginprl,ZvyaginSchlottmann,Zvyagincondmat}.
The behavior of the persistent current, $j$ in quantum dot systems is therefore
the subject of considerable controversy.

In this letter we attempt to settle this controversy 
using exact diagonalization (ED) and Density Matrix Renormalization Group (DMRG). 
The calculations have been performed using fully parallelized programs on distributed
SHARCNET facilities.
The DMRG work turns out to be very cpu-intensive due to 
the necessity of using periodic boundary conditions to obtain 
a persistent current, the complex form of 
the Hamiltonian at non-zero flux, the necessity of 
calculating a derivative of the groundstate energy to obtain the current:
$j=-edE/d\alpha$, and the fact that the current is $o(1/L)$. 
In the case of the EQD we find good agreement 
with the expected scaling picture and obtain useful results on 
the crossover functions, $f$.  For the SCQD we find 
an extremely slow crossover with varying $L$ or $J_K$ but 
we present both numerical and analytical evidence 
that $j$ scales to zero at small $\xi_K/L$ at 1/2-filling, as 
predicted by the previous RG approach. 
We show that particle-hole symmetry
breaking leads, for small $J_k/t$, to small non-universal corrections to
$jL$.  Surprisingly, these produce a small non-zero value of $j_eL$, the current for even $N$, at
$L\to \infty$ for the SCQD.

\begin{figure}
\begin{center}
\includegraphics[clip,width=8cm]{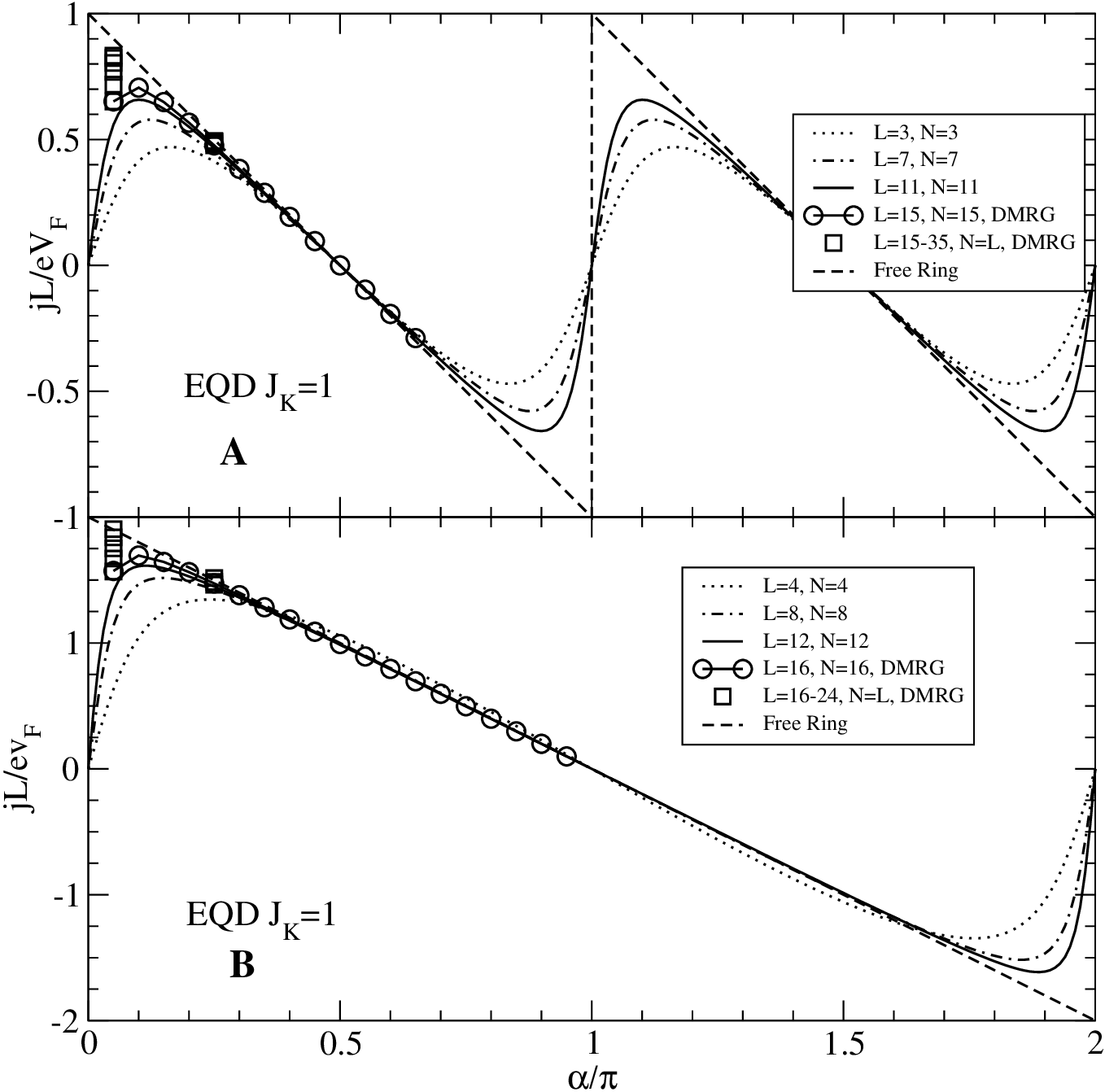}
\caption{The current at 1/2-filling for $N=4p-1$ (A) and $N=4p$ (B) for the EQD. Results
are shown for a number of system sizes with $J_K=1$ as a function of $\alpha/\pi$. 
The results for the small system sizes
are obtained using exact diagonalization methods while the larger system sizes (circles) have been
obtained using DMRG techniques.
\label{fig:HalfOdd}}
\end{center}
\end{figure}

\begin{figure}
\begin{center}
\includegraphics[clip,width=8cm]{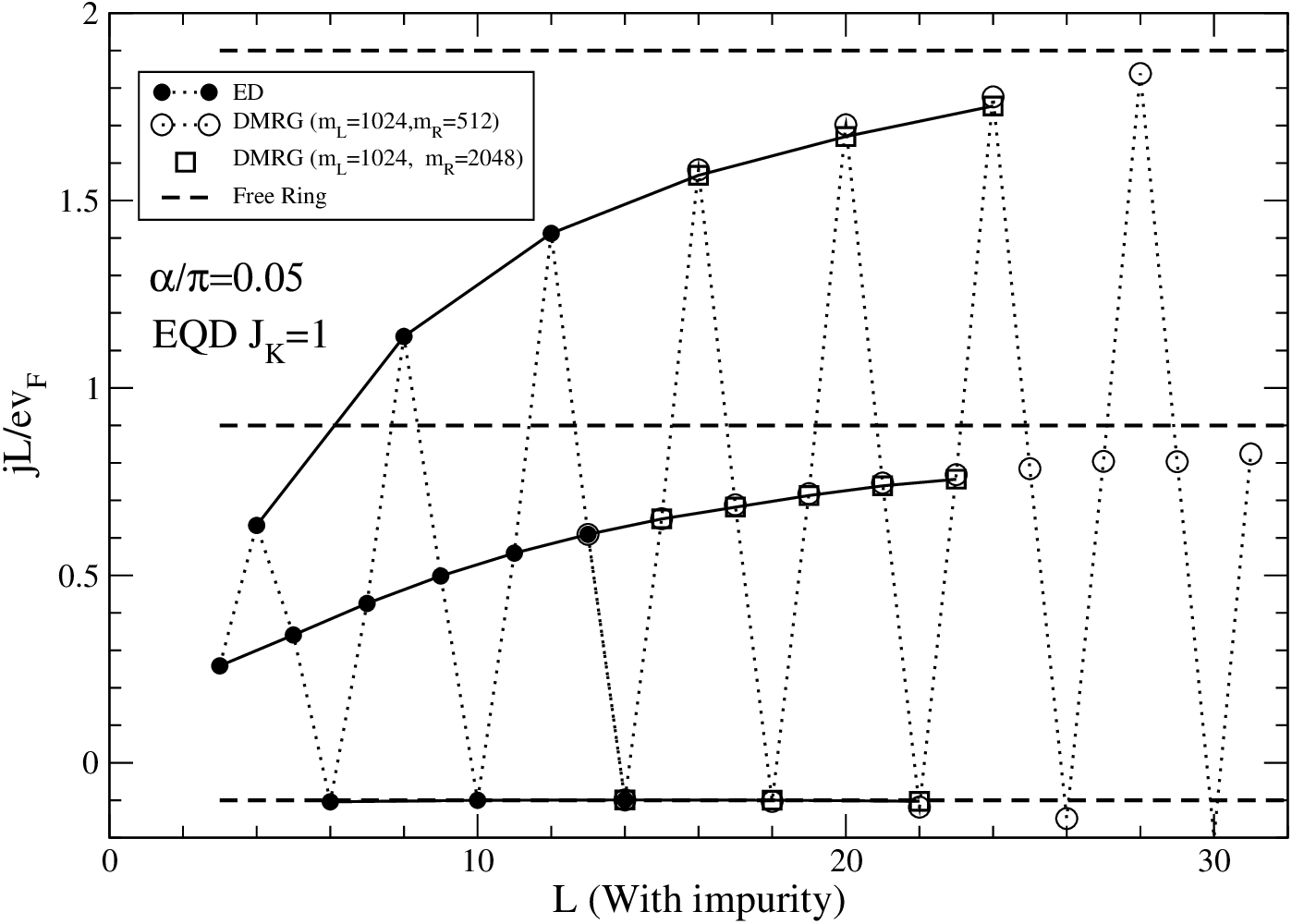}
\caption{The current $jL/ev_F$ for $J_K=1$ and $\alpha/\pi=0.05$ as a function of $L$ for the EQD.
The results for the small system sizes
are obtained using exact diagonalization methods ($\bullet$) while the larges system sizes have been
obtained using DMRG techniques with (m$_L$=1024,m$_R$=512) ($\circ$). The $\Box$ indicate DMRG results
obtained using (m$_L$=1024, m$_R$=2048). 
The dashed lines indicate the free ring result.
\label{fig:jL.a05.J=1.00}}
\end{center}
\end{figure}

\begin{figure}
\begin{center}
\includegraphics[clip,width=8cm]{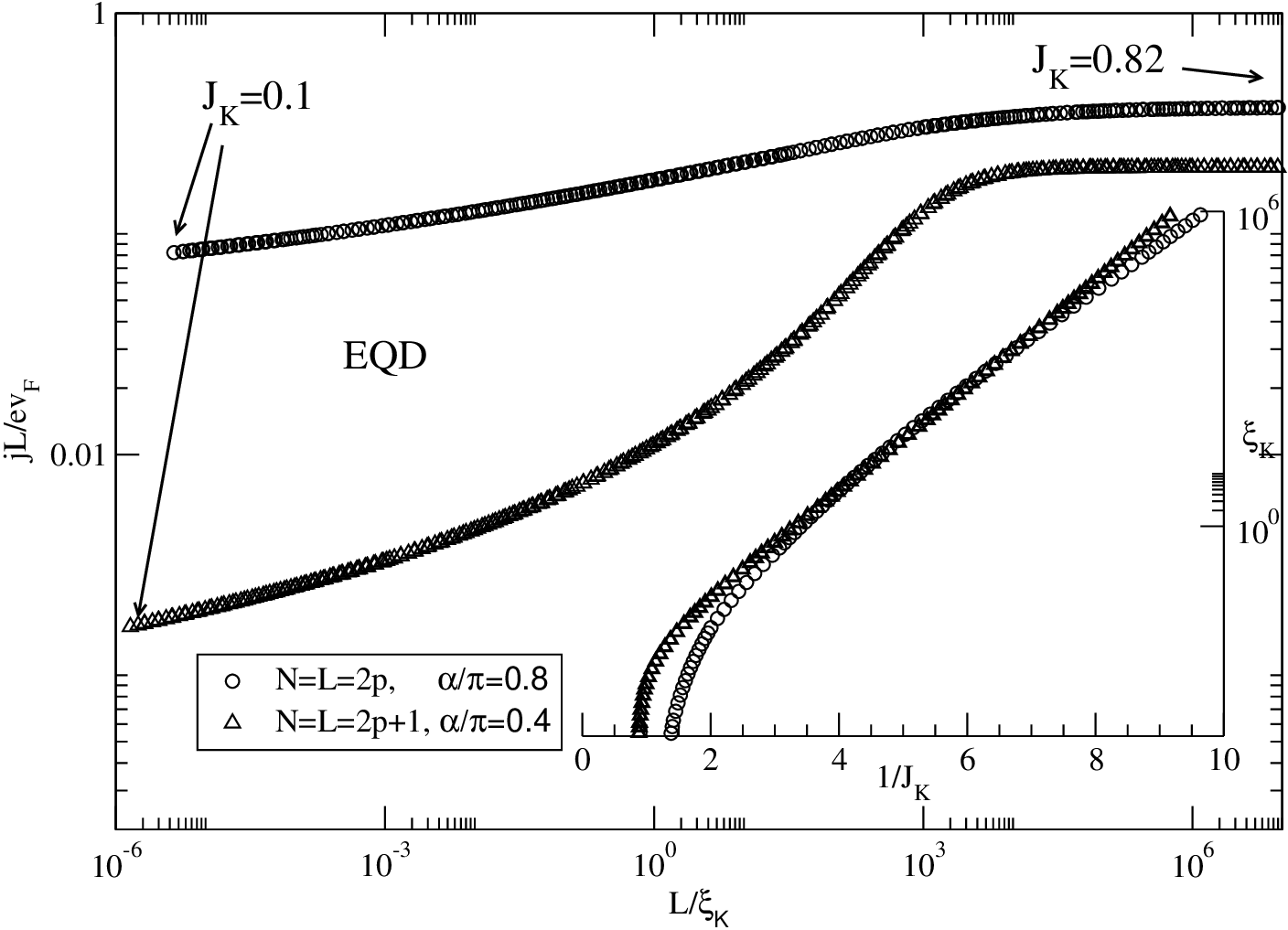}
\caption{The current at 1/2-filling $jL/ev_F$, at $\alpha/\pi=0.4$ ($N=4p$) $\alpha/\pi=0.8$ ($N$ odd), plotted vs. $L/\xi_K$ obtained 
by combining our results for various values of $J_K$ for the EQD. We fix $\xi_K(J_K=0.3)\equiv 1$ and obtain
the remaining $\xi_K$ relatively to $\xi_K(J_K=0.3)$ by rescaling. The inset shows the obtained $\xi_K(J_K)$ as a function of
$1/J_K$.
\label{fig:ScaleEQD}}
\end{center}
\end{figure}

{\it Embedded Quantum Dot}: In Fig. (\ref{fig:HalfOdd}) we 
plot $j$ vs. $\alpha/\pi$ for a fixed $J_K=1$, and 
various $L=N$.  Here both $L$ and $N$ includes contribution from the impurity site.
Note that a crossover is seen between the weak coupling behavior at smaller 
lengths  (small and sinusoidal) to strong coupling ideal ring
behavior at the largest lengths (larger and the saw-tooth). To further
illustrate this crossover, we focus on one value of the flux, $\alpha /\pi  =0.05$. In 
Fig. (\ref{fig:jL.a05.J=1.00}) we plot $jL/ev_F$ vs. $L$ for 
this fixed value of the flux for $L=N$ both even or odd. Small numerical errors in 
the DMRG calculations with m=(1024,512) are visible beyond L=24 when compared to results with m=(1024,2048).
As can be seen from Figs.~(\ref{fig:HalfOdd}) and
(\ref{fig:jL.a05.J=1.00}) the current depends strongly on whether $L=N$ is even or odd. For $L=N$ even there
is a difference between $N=4p$ and $N=4p+2$, visible in Fig~(\ref{fig:jL.a05.J=1.00}), that can be
absorbed into a re-definition~\cite{SAprb} of the flux $\tilde\alpha=\alpha+\pi N/2$. From the results shown in
Figs.~(\ref{fig:HalfOdd}) and (\ref{fig:jL.a05.J=1.00}) we conclude that $jL$ {\it increase} with $L$
towards the free ring limit contradicting Refs~\onlinecite{Kang,Ferrari}.

Weak and strong coupling results~\cite{SAprb}, and Fig. (\ref{fig:HalfOdd})
indicate that $j(\alpha )$ has period $2\pi$ for $N=L$ even 
but period $\pi$ for $N=L$ odd. The latter result follows rigorously 
at 1/2-filling from particle-hole (P-H) symmetry which takes $\alpha \to \pi -\alpha$ 
for $N=L$ odd together with time reversal which takes $\alpha \to -\alpha$.
Away from 1/2-filling, when P-H symmetry is broken, we 
might expect that the period of $j(\alpha )$ would be enlarged to $2\pi$.
This can be checked at both weak and strong coupling. In the 
weak coupling limit, in addition to the terms calculated previously~\cite{SAprb}, 
we find an additional term, of period $2\pi$ in the current for 
$N$ odd, only present away 
from 1/2-filling ($N/L\neq 1$):
\begin{equation}
\delta j_o\approx 
\frac{3J_K^2e}{\pi L}\sin \tilde\alpha \sin^2(\pi N/2L)
\frac{2\pi\cos(\frac{\pi N}{2L})}{4t}, 
\label{eq:jsin}
\end{equation}
where $\tilde\alpha=\alpha+\pi(N-1)/2$.
In the weak coupling limit, the period $2\pi$ term in $j$ for odd $N$ 
can be understood, in the continuum limit formulation with a 
linearized dispersion relation and a wave-vector cut-off which 
is symmetric around the Fermi surface, as arising from particle-hole
symmetry breaking potential scattering terms in the effective 
Hamiltonian, of $O(J_K^2)$. These terms are strictly marginal under renormalization 
group transformations and hence their effects do not grow 
larger with growing $L/\xi_K$, but remain of $O(J_K^2)$, 
where $J_K$ is the {\it bare} coupling, at all length scales. 
Thus the period $2\pi$ term in $j$ becomes negligible at 
all length scales, when $L\gg\xi_K$, in the limit of small 
bare coupling and hence do not appear in the universal 
scaling functions, $f(\xi_K/L,\alpha )$. 

We have also attempted to numerically calculate an approximation to the
scaling function $f$ by rescaling ED results out to $L=13$ for a range of $J_K$. 
Our results are shown in Fig. (\ref{fig:ScaleEQD}) where $jL/ev_F$ is plotted
vs. $L/\xi_K$ for a fixed $\alpha /\pi =0.40(0.80)$ for $L=N$ odd(even). If $\xi(J_K^0)$
is fixed at a given $J_K^0$, $\xi(J_K^1)$ can be estimated by rescaling $L/\xi_K$ until $jL/ev_F$, calculated
with $J_K^1$ superposes $jL/ev_F$ calculated with $J_K^0$ and so forth. The reults of 
Fig. (\ref{fig:ScaleEQD}) nicely confirms the scaling picture and shows that $f$ is
an {\it increasing} function of $L/\xi_K$. 

\begin{figure}
\includegraphics[clip,width=8cm]{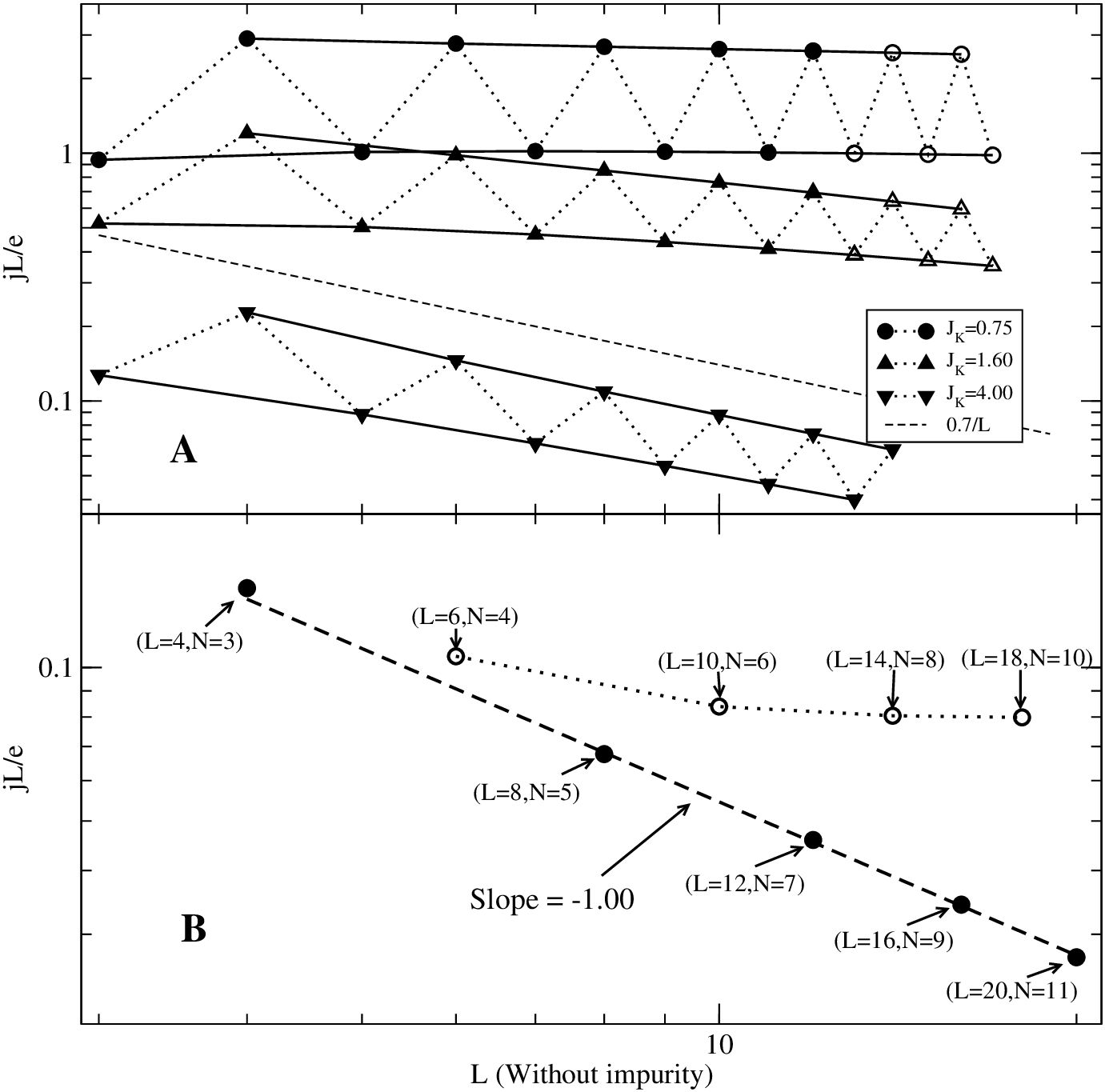}
\caption{({\bf A}) The current $jL/e$ at 1/2-filling for the SCQD for $\tilde\alpha/\pi=0.15$ and
$J_K=0.75,\ 1.60,\ 4.00$ as a function of $L$. Exact diagonalization
results (solid symbols) are shown for small system sizes and compared to DMRG results with (m$_L$=1024,m$_R$=2048) for 
$L\ge 13$ (shaded symbols). For comparison a function with pure $1/L$ behavior is shown.
Note that the current goes to {\it zero} in the strong coupling limit.
({\bf B}) The current $jL/e$ at  1/4-filling
for the SCQD for $\tilde\alpha/\pi=0.75$ and
$J_K=4.00$ as a function of $L$. Exact diagonalization
results are shown for system even $N$ ($\circ$) up to $N=10(L=18)$
and for odd $N$ ($\bullet$) up to $N=11 (L=20)$. The dashed line is a power-law fit to
the points with $N=5-11$, of slope=-1.00.
\label{fig:SCQD}}
\end{figure}

{\it Side Coupled Quantum Dot}: 
[Now $L$ does {\it not} include
the impurity site while $N$ {\it does} include the impurity electron.
For N odd $\tilde\alpha = \alpha+\pi(N-1)/2$ and for N even $\tilde\alpha = \alpha+\pi N/2$.]
In Fig. (\ref{fig:SCQD}A) we show exact diagonalization and DMRG 
results for the SCQD at $\tilde\alpha=0.15\pi$ analogous to Fig~(\ref{fig:jL.a05.J=1.00}).
These results are consistent with 
$jL$ going to zero at $L\gg\xi_K$ however it appears necessary 
to go to extremely large $L/\xi_K$ to see this behavior. 
A calculation of the scaling function, $f$, equivalent to Fig. (\ref{fig:ScaleEQD}), shows
that $f$ for the SCQD is a {\it decreasing} function of $L/\xi_K$ in contradiction to the
results of Refs.~\onlinecite{Eckle,Cho}.

In order to test our conjectured scaling behavior, we have 
studied both analytically and numerically the strong coupling 
limit, $J_K\gg1$ in greater detail. In this limit one 
electron sits at site $0$ and forms a singlet with the 
impurity spin, effectively cutting the ring at the origin. 
Perturbation theory in $1/J_K$ generates terms 
in a low energy effective Hamiltonian which couple 
the two sides of the quantum dot and thus allow for 
a small non-zero current.  By doing perturbation theory 
to 3$^{\hbox{rd}}$ order in $1/J_K$, we obtain an effective Hamiltonian
valid at $J_K\gg 1$:
\begin{equation}
H_{eff} = -\sum_{j=1}^{L-2}\left(c^{\dagger}_jc_{j+1}+{\rm h.c.}\right)+
H_T,
\label{eq:H_T}
\end{equation}
where the tunneling terms are:
\begin{eqnarray}
H_T&=&\frac{4}{9J_K^2}e^{-i\alpha}(c_1^\dagger c_{L-2}+c_2^\dagger c_{L-1})+h.c.
\nonumber \\ &&
-\frac{32}{J_K^3}e^{-2i\alpha}c^\dagger_{1\uparrow}c^\dagger_{1\downarrow}
c_{L-1\uparrow}c_{L-1\downarrow}+h.c.
\nonumber \\
&&+\frac{4}{3J_K^3}(n_1+n_{L-1}-2)e^{-i\alpha}
c_1^\dagger c_{L-1}+h.c.
\end{eqnarray}
Here $n_i$ is the electron number on site $i$. We have ignored 
additional terms in the effective Hamiltonian which do not 
contribute to the current at low orders in $1/J_K$. We may 
now evaluate the current, up to $O(1/J_K^3)$, by calculating 
the $\alpha$-dependence of the groundstate energy in first  
order perturbation theory in $H_T$. Considering 1/2-filling, 
this gives, for odd or even $N$, 
\begin{eqnarray}
j_oL/e&\approx &\frac{32}{9J_K^2}[
\tan (\frac{\pi}{2L})+
\tan (\frac{3\pi}{2L})
]\sin \tilde \alpha 
\nonumber \\
&&+\frac{128}{3J_K^3L}[2\sin \tilde \alpha -\sin (2\tilde \alpha )]
\nonumber \\
j_eL/e&\approx &\frac{32}{3J_K^3L}[1+1/\cos (\pi /L)]^2\sin 2\tilde\alpha .
\label{eq:jlarge}
\end{eqnarray}
The absence of period $2\pi$ terms for $N$ even can be 
understood from the  P-H symmetry of the unperturbed 
Hamiltonian in Eq. (\ref{eq:H_T}) which is broken, for $N$ even,  
by $H_T$.  Note that these formulas predict that 
$j$ goes to zero as $1/L^2$ at large $L$. In the strong 
coupling limit, this 
amounts to an analytic proof that the current 
is not the same as for the ideal ring, in contradiction 
with the claims of Ref.~\onlinecite{Eckle}.
We have 
verified that Eqs. (\ref{eq:jlarge}) are in excellent 
agreement with our numerical results for large $J_K$. 
However, we find that surprisingly large values of $J_K$, of 
about 100, 
are necessary before these formulas become accurate.
This is consistent with the results of our study 
of length dependence of $j$ which suggests a very 
slow cross over from weak to strong coupling 
behavior as $L/\xi_K$ is varied. 

From Eqs.~(\ref{eq:jlarge}) it is also possible to calculate
the current away from 1/2-filling, at large
$J_K$ we find:
\begin{eqnarray}
\lefteqn{j_oL/e\approx\frac{32}{9J_K^2}\sin (\tilde \alpha )}&&\nonumber\\
&&\times\left[\sin \frac{\pi (N-1)}{2L}\tan \frac{\pi}{2L}
-\sin \frac{3\pi (N-1)}{2L}\tan \frac{3\pi}{2L}\right]\nonumber\\
\lefteqn{j_eL/e \approx  \frac{32}{9J_K^2} \sin (\tilde \alpha )
\left[ \frac{\cos (\frac{\pi (N-1)}{2L})}{\cos (\frac{\pi }{2L})}-
\frac{\cos (\frac{3\pi (N-1)}{2L})}{\cos (\frac{3\pi}{2L})}\right]}&&
\label{eq:scqdoff}
\end{eqnarray}
Note that $j_oL/e$ is $O(1/L)$ and actually gets smaller as we move away from
1/2-filling. We see that $j_oL$ {\it approaches zero} as $L\to\infty$.
Surprisingly, we see that $j_eL$ attains a non-zero limit as $L\to
\infty$, away from 1/2-filling, in stark contrast to $j_oL$
and the behavior of both $j_oL$ and $j_eL$ at 1/2-filling.
We have numerically verified Eqs.~(\ref{eq:scqdoff}) in detail finding excellent
agreement for $J_K>100$. For an intermediate coupling of $J_K=4$ we show
ED results for $jL/e$ in Fig~(\ref{fig:SCQD}B), at a flux $\tilde\alpha/\pi=0.75$ and 1/4-filling, clearly
approaching a non-zero limit for $N$ even 
with opposite sign and a much smaller amplitude than that of the ideal ring.
For odd $N$, $jL/e\sim {\rm const}/L$ in accordance with
Eq.~(\ref{eq:scqdoff}).
This O($1/L$) behavior of $j_e$ away from 1/2-filling results from
non-universal particle-hole symmetry breaking terms in the low energy
effective Hamiltonian.  For small $J_K/D$ (where $D\approx t$ is the
bandwidth) these terms are small, and remain so under renormalization.
They contribute to $j_e$ at $L\gg\xi_K$, which has the form:
$j_e=(ev_F/L)[(A\xi_K/L)\sin 2\alpha + B(J_K/D)^2\sin \tilde \alpha ]$
where $A$ is a universal constant and $B$ is a non-universal constant,
both of O(1). Thus the universal ($1/L^2$) behavior is destroyed at large
length scales, $\approx \xi_K(D/J_K)^2\gg\xi_K$, away from 1/2-filling, in
contrast to the claim in Ref. (\onlinecite{SAprb}).  Note however that
the current at these large length scales is smaller by a factor of
$(J_K/D)^2$ than that of an ideal ring, in contrast to the claim in Ref.
(\onlinecite{Eckle}).

Following Nozi\`eres\cite{Nozieres} we have developed a local Fermi liquid
theory description of the low temperature fixed point of the SCQD.  This
fixed point corresponds to the impurity forming a singlet with one
conduction electron and the remaining low energy electrons being repelled
from the origin, corresponding to a broken ring with zero persistent
current. In the 1/2-filled case, the leading irrelevant operators at this
fixed point, which control the persistent current at large $L$, are
$(\psi^\dagger_+\vec \sigma \psi_-)^2$, $\psi^\dagger_+\vec \sigma \psi_-
\cdot (\psi^\dagger_+\vec \sigma \psi_+ +\psi^\dagger_-\vec \sigma
\psi_-)$ and their complex conjugates. Here $+$ and $-$ label the electron
operators on the right and left hand side of the impurity.  In the
limit of weak bare Kondo coupling, or large $\xi_K$, the coupling
constants in front of these terms can all be fixed uniquely up to one
overall factor with dimensions of inverse energy, proportional to the
inverse of the Kondo temperature. This theory predicts that the persistent
current scales to zero as $ev_F^2/T_KL^2$ at 1/2-filling, or for even $N$
at arbitrary filling, as well as making various other predictions that
could be compared with numerical simulations and
experiments.\cite{LongKondo}

In conclusion, we have given strong numerical evidence and analytical results
in support of the
scaling picture of the persistent current in quantum dot systems resolving
a number of outstanding controversies.
This research is supported by NSERC of Canada, CFI, SHARCNET and CIAR.
IA acknowledges interesting conversations with P. Simon.
\bibliography{persistent}

\end{document}